\documentclass[twocolumn,aps,jcp,amsmath,amssymb,superscriptaddress,showpacs]{revtex4-1}

\usepackage{graphicx}
\usepackage{graphics}
\usepackage{dcolumn}
\usepackage{bm}
\usepackage{color}
\usepackage{soul}
\usepackage{textcomp}
\usepackage{soul}
\usepackage{rotating}
\usepackage{setspace}
\usepackage[mathlines]{lineno}
\usepackage{textcomp}
\usepackage{mathrsfs}
\usepackage{amssymb}
\usepackage{mathcomp}
\def\prjctr#1#2{\left|#1\left\rangle \right\langle#2\right| }

\def\ket#1{|\,#1\, \rangle}

\usepackage{microtype}
\usepackage[breaklinks=true,hyperindex=true,
pdftitle={EExact Factorization of the Time-Dependent Electron-Nuclear Density Operator},
colorlinks=true,pagebackref=false,citecolor=blue,plainpages=false,pdfpagelabels,
linkcolor=blue,urlcolor=blue]{hyperref}

\def\eu{{\rm e}}
\def\iu{{\rm i}}

\begin{document}

\newcommand{\affA}{Department of Chemistry and Chemical Biology, Harvard University, Cambridge, MA 02138 USA.}

\title{
Quantum Process Tomography by 2D Fluorescence Spectroscopy
}

\author{Leonardo A. Pach\'on}
\affiliation{Grupo de F\'isica At\'omica y Molecular, Instituto de F\'{\i}sica,  Facultad de Ciencias Exactas y Naturales, 
Universidad de Antioquia UdeA; Calle 70 No. 52-21, Medell\'in, Colombia}
\affiliation{\affA}

\author{Andrew H. Marcus}
\affiliation{Department of Chemistry and Biochemistry, Oregon Center for Optics, Institute of Molecular Biology, 
University of Oregon, Eugene, Oregon 97403, United States}

\author{Al\'an Aspuru-Guzik}
\affiliation{\affA}

\begin{abstract}
Reconstruction of the dynamics (quantum process tomography) of the single-exciton manifold in 
energy transfer systems is proposed here on the basis of two-dimensional fluorescence spectroscopy 
(2D-FS) with phase-modulation.
The quantum-process-tomography protocol introduced here benefits from, e.g., the sensitivity 
enhancement ascribed to 2D-FS.
Although the isotropically averaged spectroscopic signals depend on the quantum yield parameter 
$\Gamma$ of the doubly-excited-exciton manifold, it is shown that the reconstruction of the dynamics 
is insensitive to this parameter.
Applications to foundational and applied problems, as well as further extensions, are discussed.
\end{abstract}

\date{\today}

\pacs{03.65.Yz, 05.70.Ln, 37.10.Jk}

\maketitle

\section{Introduction}
Since Chuang and Nielsen's 1996 seminal proposal to experimentally 
reconstruct the evolution operator of a quantum black box \cite{CN96}, a variety of experiments 
have been proposed \cite{ML06,BPP08,SK&11} and some of these have been implemented 
\cite{WZ&07,SLP10,NB&13}.
Although most of these ideas involved relatively ``clean'' optical systems in 
the context of quantum information processing (QIP), recently there has been growing interest in 
the application of quantum process tomography to study electronically coupled molecular systems 
\cite{YK&11,YA11,YA&13}.
Such combined theoretical and experimental studies on excitation energy transfer 
serves to bring together the QIP and physical chemistry communities.


The current version of QPT for excitonic systems \cite{YK&11,YA11,YK&14} is based on the method 
of 2D Photon Echo Spectroscopy (2D-PES) \cite{Muk99,Cho10}.
It therefore relies on the wave-vector phase-matching condition, which works for macroscopic
systems with many chromophores  (see Refs.~\cite{Muk99,Cho10,YK&14}  for details).
For this reason, the proposal developed in Refs.~\cite{YK&11,YA11,YA&13} is not suitable for 
single-molecule QPT, and it is hence desirable to implement the phase-cycling (PCT) or 
phase-modulation techniques (PMTs) \cite{TDM06,TLM07}.
In most 2D-PES experiments, the system interacts with three non-collinear ultrafast laser pulses, 
which gives rise to a third-order polarization that propagates in the wave-vector matched direction 
\cite{Muk99,Cho10,YK&14}. 
The transmitted signal must be separated from background laser light, which is inadvertently scattered 
by the sample in the same direction as the third-order signal. 
The presence of background scattering is a limiting factor to the sensitivity of 2D-PES experiments. 
In 2D-FS, the system interacts with four collinear laser pulses, and the signal is detected by monitoring 
nonlinear contributions to the ensuing fluorescence signal \cite{TLM07,PW&12}. 
The red-shifted fluorescence can be easily separated from background scattered laser light by using 
long-pass spectral filters. Specific nonlinear contributions to the fluorescence signal are isolated according 
to the phase-modulation schemes described in previous work \cite{TLM07,PW&12}. 
Because the 2D-FS method is based on the detection of incoherent fluorescence signals, it may be 
applied to systems of small numbers of chromophores, quantum dots, or thin film materials.

Despite the high-sensitivity advantages afforded to fluorescence-detected PMTs, which are useful for 
studies of biological molecules and molecular aggregates \cite{VJM13,WJ&13,WL&13}, these methods 
have been less commonly practiced than four-wave mixing approaches to 2D-PES.
However, recent theoretical \cite{TLM07} and experimental \cite{LP&11,PW&12} progress with classical 
light have enabled PMTs for a variety of complex molecular systems relevant to exciton dynamics.
These recent developments, and the general theory of open quantum systems-- i.e., quantum systems 
coupled to the environment-- are combined here to formulate a self-consistent theory of QPT that is 
based on collinear PMT with synchronous detection.

For the ideal situation when nonradiative processes are neglected in the doubly-excited-exciton manifold,
the quantum yield parameter of this manifold is set to $\Gamma=2$.
Under this circumstance, it was shown that 2D-FS coincides with  2D-PES \cite{TLM07,LP&11,PW&12}.
It is shown below that this equivalence also holds at the level of quantum process tomography, i.e.,
the protocol introduced here generalizes the protocol in Refs.~\cite{YK&11,YA11} to the 
more realistic situation $0\le\Gamma<2$.
%

\section{Initial considerations on QPT, system model and 2D-FS}
Before introducing the reconstruction of the dynamics, it is necessary to state some remarks
on the basics of process tomography, the system-of-interest model and 2D-FS.

\textit{Quantum Process Tomography Tensor}\textemdash In quantum mechanics, the state of a 
physicochemical system S is described by a density operator $\hat{\rho}$. 
Time evolution of quantum states is governed by the Schr\"odinger equation, which is 
linear in the state of the system.
This linearity allows for a description of the system's dynamics in terms of a linear map, 
$\hat{\chi}_t: \hat{\rho}_0 \mapsto \hat{\rho}_t$.
After projecting onto a complete orthonormal basis $\{|n\rangle\}$, the map reads 
\begin{equation}
\label{equ:defXpqrs}
\langle n | \hat{\rho}(t) | m  \rangle = \sum_{\mu\nu} \chi_{nm\nu\mu}(t)
 \langle \nu | \hat{\rho}(0) | \mu  \rangle,
\end{equation}
where $\chi_{nm\nu\mu}(t)$ stands for the process tomography tensor \cite{Cho75,YK&11,ML06}.
For Hamiltonian dynamics with $\hat{H} |n\rangle = E_n |n\rangle$, 
$\chi_{nm,\nu\mu}(t) = \mathrm{e}^{-\mathrm{i} (E_m - E_n)t/\hbar}
\delta_{n\nu}\delta_{m\mu}$  \cite{PB13a,PB13b,PB14}.
Thus, population-to-coherence [$\chi_{nm\nu\nu}(t)$] and the reverse 
$[\chi_{nn\nu\mu}(t)]$ process are prevented by the Kronecker deltas $\delta_{n\nu}\delta_{m\mu}$.
Clearly, this restriction is not present if driving fields are present or if the system of interest
is coupled to its environment  \cite{PB13a,PB13b,PB14}.

In the general case of open quantum systems, the functional form of Eq.~(\ref{equ:defXpqrs}) remains
valid under some conditions. 
(i) If the coupling to the bath is weak, Eq.~(\ref{equ:defXpqrs}) holds for Markovian and non-Markovian 
processes and the process tensor is independent of the initial state (see, e.g., Refs.~\cite{YK&11,YA11} 
and references therein).
(ii) If the coupling to the bath is strong, and the initial system-environment correlations cannot be neglected,
Eq.~(\ref{equ:defXpqrs}) holds after including those initial correlations in $\chi_{nm\nu\mu}(t)$ (see 
Refs.~\cite{PB13a,PB13b,PB14} for details).
(iii) Because the initial bath correlations vanish at high temperature, even for strong coupling \cite{PT&14}, 
then $\chi_{nm\nu\mu}(t)$ can be defined independently of the initial state in the strong coupling
regime entered at high temperatures.

After identifying the conditions under which Eq.~(\ref{equ:defXpqrs}) holds, it is relevant to consider 
some of the main properties of the QPT tensor \cite{YK&11}, namely,
\begin{align}
\label{equ:prop1}
\chi_{nm\nu\mu} = \chi_{mn\mu\nu}^*,
\\
\label{equ:prop2}
\sum_{n} \chi_{nn\mu\nu}(T) = \delta_{\mu\nu}, 
\\
\label{equ:prop3}
\sum_{nm\nu\mu} z^*_{n\nu} \chi_{nm\nu\mu} z_{m\mu} \ge 0,
\end{align}
where $z$ is any complex valued vector. 
Equation~(\ref{equ:prop1}) ensures the Hermitian character of the density operator, 
$\hat{\rho} = \hat{\rho}^\dag$, while Eq.~(\ref{equ:prop2}) guaranties probability conservation, 
$\mathrm{tr}\hat{\rho}(t)=1$.
The last property is a consequence of the fact that $\hat{\rho}(t)$ remains positive-semidefinite
under unitary operations.

The objective of QPT is the experimental reconstruction of the process tomography tensor 
$\chi_{nm\nu\mu}(t)$.

\textit{Model}\textemdash
Consider an excitonic dimer described by $\hat{H}_{\mathrm{S}}$ and given by
\begin{equation}
\label{equ:DmrHamiltonian}
\hat{H}_{\mathrm{S}} = \omega_{1} \hat{a}_{1}^\dag \hat{a}_{1}
+\omega_{2} \hat{a}_{2}^\dag \hat{a}_{2}
+J\left( \hat{a}_{1}^\dag \hat{a}_{2} + \hat{a}_{2}^\dag \hat{a}_{1} \right),
\end{equation}
where $ \hat{a}_{i}^\dag$ and  $\hat{a}_{i}$ are the creation and annihilation operators for site 
$i$, $\varpi_1\neq\varpi_2$ are the site energies while $J\neq0$ is the Coulombic coupling
between chromophores.
By defining the average frequency $\varpi=\frac{1}{2}(\varpi_1+\varpi_2)$,
the half-difference $\Delta = \frac{1}{2}(\varpi_1-\varpi_2)$ and the mixing angle 
$\theta=\frac{1}{2}\arctan(J/\Delta)$, it is possible to introduce the creation and annihilation operators,
$\hat{c}_{p} = \cos\theta \hat{a}_1 + \sin\theta \hat{a}_2$ and 
$\hat{c}^\dag_{p} = \sin\theta \hat{a}^\dag_1 + \cos\theta \hat{a}^\dag_2$,
of the $p$-th delocalized exciton state with energy
$\varpi_{p}=\varpi \pm \Delta \sec 2\theta$ and $p \in \{e,e'\}$.
Starting from the ground state $\ket{g}$, the single-exciton states are conveniently defined as
$\ket{e} = \hat{c}^\dag_{e}\ket{g}$ and $\ket{e'} = \hat{c}^\dag_{e'}\ket{g}$, while the biexciton 
state as $\ket{f} = \hat{a}^\dag_{1} \hat{a}^\dag_{2}\ket{g} = \hat{c}^\dag_{e} \hat{c}^\dag_{e'}\ket{g}$
with $\varpi_f = \varpi_1+\varpi_2 = \varpi_e + \varpi_{e'}$.
The dipole vectors at each site are set to $\mathbf{d}_1 = d_1 \mathbf{e}_z$ and 
$\mathbf{d}_2 = d_2 \cos(\phi) \mathbf{e}_z + d_2 \sin(\phi) \mathbf{e}_x$.
So that, $\boldsymbol{\mu}_{eg} = d_2 \sin\theta \sin\phi \, \mathbf{e}_x 
+ \left( d_1 \cos\theta + d_2 \sin\theta \cos\phi \right) \mathbf{e}_z$, 
 $\boldsymbol{\mu}_{e^\prime g} = d_2 \cos\theta \sin\phi \,\mathbf{e}_x 
+ \left(- d_1 \sin\theta + d_2 \cos\theta \cos\phi \right) \mathbf{e}_z$,
 $\boldsymbol{\mu}_{fe} = d_2 \cos\theta \sin\phi \, \mathbf{e}_x 
+ \left(d_1 \sin\theta + d_2 \cos\theta \cos\phi \right) \mathbf{e}_z$ and 
 $\boldsymbol{\mu}_{fe^\prime} = -d_2 \sin\theta \sin\phi \, \mathbf{e}_x 
+ \left(d_1 \cos\theta - d_2 \sin\theta \cos\phi \right) \mathbf{e}_z$.

Although exciton-exciton binding or repulsion terms are not included here, it is considered 
that each excitonic manifold contributes to the spectroscopic signal with a weight given by their 
fluorescence quantum yield coefficients $\Gamma_\nu$.
Specifically, it is assumed that the quantum yield of the two singly excitonic states are the same
and equal to 1, while for the doubly excitonic manifold, it is assumed that $\Gamma_f = \Gamma$
with $0\le \Gamma \le2$.
In the ideal case in which two photons are emitted via the path 
$\ket{f} \rightarrow \ket{e,e'} \rightarrow \ket{g}$, $\Gamma =2$.
It is possible that singlet-singlet annihilation would convert a doubly-excited excition 
into a singly-excited exciton \cite{BH&01}, which in the absence of non-radiative decay would 
result in $\Gamma = 1$.  
However, because of the abundance of non-radiative relaxation pathways for highly excited 
states, the quantum yield of the doubly-excitonic manifold is expected to be smaller than that 
of the singly excitonic manifold, so that values smaller than unity are expected. 
For example, for membrane-supported self-assembled porphyrin dimers, it was found that 
$\Gamma = 0.31$ \cite{LP&11,PW&12}.

For convenience, the dimer Hamiltonian can be written as 
$\hat{H}_{\mathrm{S}} = \sum_{\nu=\{g,e,e^\prime,f\}} \omega_\nu \prjctr{\nu}{\nu}$. 
To account for the influence of the local vibrational environment in the excitonic dimer, coupling to 
a thermally equilibrated phonon bath at inverse temperature $\beta$ is considered next.
Specifically, the Hamiltonian of the environment is given by 
$\hat{H}_{\mathrm{E}} = \sum_{p=e,e^\prime} \sum_n  \Omega_{n,p} 
\left(\hat{b}_{n,p}^\dag \hat{b}_{n,p} + 1/2 \right)$, where $ \Omega_{n,p}$ denotes the frequency
of the environment modes. 
The interaction is described by 
$\hat{H}_{\mathrm{SE}} = \hat{E}_{e} \prjctr{e}{e} +  \hat{E}_{e^\prime} \prjctr{e^\prime}{e^\prime}
+\left(\hat{E}_{e}  + \hat{E}_{e^\prime} \right) \prjctr{e^\prime}{e^\prime}$ with
$\hat{E}_{p} = \sum_n \lambda_{n,p} \left( \hat{b}_{n,p}^\dag + \hat{b}_{n,p} \right)$.
$\hat{b}_{n,p}^\dag$ and $\hat{b}_{n,p}$ are the creation and annihilation bosonic operators of 
the $n-$th mode of the vibrational environment in the $p-$site. $\lambda_{n,p}$ measures the 
interaction strength between the $n-$th mode of the environment and the $p-$th site.
The net effect of the local environment is encoded in the spectral density 
$J_n = \sum_n  \Omega_{n,p}^2  \lambda_{n,p}^2  \delta(\omega-\omega_n)$.

\textit{2D Fluorescence Spectroscopy (2D-FS)}\textemdash
The main difference between the QPT scheme introduced below and previous QPT proposals 
is the spectroscopic technique, 2D-FS, which the present proposal is based on.
It is therefore relevant to discuss the main differences and advantages that 2D-FS has over, e.g.,
2D-PES.
The 2D-FS method isolates the nonlinear optical response of a material system by monitoring 
fluorescence signals. 
Because fluorescence can be efficiently separated from background scattered light, the 2D-FS 
approach can be used to perform experiments that require very high detection sensitivity 
\cite{TLM07,LP&11,PW&12}.  
%
%
Moreover, the collinear beam geometry used in 2D-FS has the advantage that every illuminated 
molecule experiences the same optical phase condition at every instant in time. 
Since the incoherent fluorescence signal is emitted isotropically, very small numbers of molecules 
may be studied in this way.  \cite{TLM07,LP&11,PW&12}.

The 2D-FS observable is proportional to the fourth-order excited populations, 
\begin{equation}
\langle \hat{A}(t) \rangle = \mathrm{tr} \hat{A} \hat{\rho}^{(4)}(t),
\end{equation}
with $\hat{A}=\sum_{\nu=\{e,e',f\}} \Gamma_\nu \prjctr{\nu}{\nu}$, 
generated by the action of the operator $\hat{V}(t')$ that comprises the 
excitation by four weak non-overlapping laser pulses,
\begin{equation}
\hat{V}(t') = -\lambda \sum_{i=1}^4 \hat{\boldsymbol{\mu}}\cdot \mathbf{e}_i
E(t'-t) \left[ \mathrm{e}^{-\mathrm{i}\omega_i (t'  - t_i) + \phi_i} + c.c. \right].
\end{equation}
Here $\lambda$ denotes the maximum intensity of the pulses' electric field, and $\hat{\boldsymbol{\mu}}$ 
the dipole operator. $\mathbf{e}_i$, $t_i$,  $\omega_i$ and $\phi_i$ stand for the polarization 
vector, time center, frequency and phase of the $i$-th laser pulse.
The pulse envelope $E(t)$ is chosen to be Gaussian with fixed width $\sigma$, i.e., 
$E(t) = \mathrm{e}^{-t^2/2\sigma^2}$.
In the model under consideration, the only optically allowed transitions are between states
differing by one excitation.
Hence, the only non-vanishing dipole transition matrix elements are
$\boldsymbol{\mu}_{ij}=\boldsymbol{\mu}_{ji}$ with $ij=\{eg,e'g,fe,fe'\}$.
Details about the derivation and the explicit functional form the fourth-order density matrix 
can be found in Appendices  \ref{app:InitalPrep} and \ref{app:FinalPrep}.

For the purpose  of extracting the QPT tensor from the 2D-FS experimental signals, only the 
rephasing signals with global phase $\phi_{\mathrm{reph}} = -\phi_1 + \phi_2 + \phi_3 -\phi_4$ 
will be considered below (see Fig.~\ref{fig:Feynman2DFS}).
Thus, assuming that the rotating wave approximation (RWA) holds, the interactions with the 
electromagnetic fields are characterized by
\begin{align}
\hat{V}_1 &= -\lambda \hat{\boldsymbol{\mu}}^< \cdot \mathbf{e}_1 E(t-t_1) 
\mathrm{e}^{\mathrm{i} \omega_1 (t-t_1)},
\\ 
\hat{V}_2 &= -\lambda \hat{\boldsymbol{\mu}}^> \cdot \mathbf{e}_2 E(t-t_2) 
\mathrm{e}^{-\mathrm{i} \omega_2 (t-t_2)},
\\
\hat{V}_3 &= -\lambda \hat{\boldsymbol{\mu}}^> \cdot \mathbf{e}_3 E(t-t_3) 
\mathrm{e}^{-\mathrm{i} \omega_3 (t-t_3)},
\\
\hat{V}_4 &= -\lambda \hat{\boldsymbol{\mu}}^< \cdot \mathbf{e}_4 E(t-t_4) 
\mathrm{e}^{\mathrm{i} \omega_4 (t-t_4)},
\end{align}
where $\hat{\boldsymbol{\mu}}^< = \sum_{\omega_p<\omega_q} 
\boldsymbol{\mu}_{pq} |p\rangle \langle q|$ promotes emissions from the ket and
absorptions on the bra, and $ \hat{\boldsymbol{\mu}}^> = (\hat{\boldsymbol{\mu}}^{<})^{\dag}$
induces the opposite processes.
For this particular selection of the global phase, $\phi_{\mathrm{reph}} = -\phi_1 + \phi_2 + \phi_3 -\phi_4$, 
the 2D-FS signals are equivalent to the rephasing spectroscopic signals in the photon-echo direction 
$\mathbf{k}_{\mathrm{PE}} = -\mathbf{k}_1 + \mathbf{k}_2 + \mathbf{k}_3$ when 
$\Gamma=2$ \cite{TLM07,LP&11,PW&12}.
It is shown below that the present QPT protocol reduces to the protocol in Refs.~[\citenum{YK&11,YA11}] 
when $\Gamma=2$ as well.

\section{2D-FS QPT}
As stated above, the main goal of QPT is the reconstruction of the dynamics of the density
operator.
In doing so, it is assumed that the structural parameters of the model, namely, the transition
frequencies $\varpi_{ij} = \varpi_i - \varpi_j$ and the electric dipole transition matrix element 
$\boldsymbol{\mu}_{ij}$ are all known. 
This prerequisite is not an issue because information about the transition frequencies is routinely
obtained from linear absorption spectra, and the transition dipole directions can be inferred form 
structural measurements and polarization spectroscopy \cite{YK&11}.  

Once the structural parameters are defined, the reconstruction of the dynamics comprises three \
main parts: (i) initial state preparation, (ii) evolution and (iii) final state detection.
In describing these stages, it is useful to introduce the standard time intervals 
$\{\tau,T,t\}$ instead of the time center $t_i$ of each pulse \cite{Muk99,Cho10}. 
The time difference between the second and the first pulse defines the \emph{coherence time} 
interval $\tau=t_2-t_1$.
The time interval between the third and the second pulse, $T=t_3-t_2$ is known as the \emph{
waiting time}, which defines the quantum channel to be characterized by the QPT scheme.
Finally, the difference between the fourth and the third pulse, $t=t_4-t_3$, denotes the 
\emph{echo time}. 

\textit{Initial State Preparation}\textemdash
The excitonic system, before any electromagnetic perturbation, is assumed to be in the ground 
state, $\hat{\rho}(-\infty) = \prjctr{g}{g}$.
Thus, the basic idea is to make use of the first two pulses to prepare the effective initial density
matrix at $T=0$, $\hat{\rho}^{\omega_1,\omega_2}_{\mathbf{e_1},\mathbf{e_2}}(T=0)$, and use the
last two pulses to read out the state.

After applying second order perturbation theory in $\lambda$, and under the assumption that
the RWA holds in this case (see Appendix~\ref{app:InitalPrep} for details), the effective initial state reads
\begin{equation}
\label{equ:InitialState}
\begin{split}
\hat{\rho}^{\omega_1,\omega_2}_{\mathbf{e_1},\mathbf{e_2}}(0)
=&-\hspace{-0.125cm}\sum_{p,q\in \{e,e'\}} \hspace{-0.125cm}C^p_{\omega_1} C^q_{\omega_2} 
(\boldsymbol{\mu}_{pg}\cdot\mathbf{e}_1)(\boldsymbol{\mu}_{qg}\cdot\mathbf{e}_2)
\\
&\times \mathcal{G}_{gp}(\tau) \left(|q\rangle \langle p | - \delta_{pq} |g\rangle \langle g |  \right),
\end{split}
\end{equation}
where $\mathcal{G}_{ij}(\tau)$ is the propagator of the optical coherence $\prjctr{i}{j}$.
For simplicity, it can be assumed as 
$\mathcal{G}_{ij}(\tau) = \Theta(\tau)\exp[(-\mathrm{i} \varpi_{ij} - \Gamma_{ij})\tau]$
begin $\Gamma_{ij}$ dephasing rates, and the Heaviside function $\Theta(\tau)$ ensures
causality.
The coefficients $C^p_{\omega_i}$ are purely imaginary and given by 
$C^p_{\omega_i} = \mathrm{i}\lambda \sqrt{2\pi \sigma^2} 
\mathrm{e}^{-\sigma^2(\varpi_{pg} - \omega_i)}$. 
Because at this level there is no influence of the doubly-excited exciton manifold, the 
effective initial state in Eq.~(\ref{equ:InitialState}) coincides with the effective initial state
prepared by 2D-PES in Ref.~\cite{YK&11}.

\begin{center}
\begin{figure*}
\includegraphics[width=1.75\columnwidth]{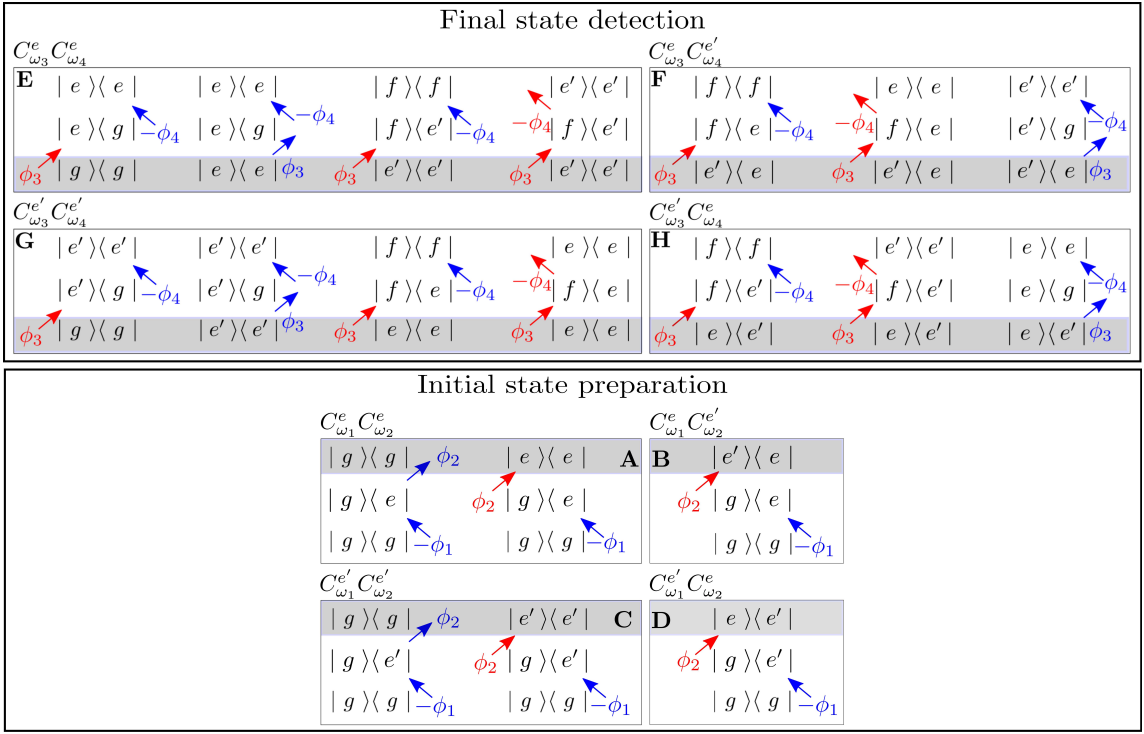}
\caption{Double-sided Feynman's diagrams for the initial state preparation (lower panel)
and state detection (upper panel) that lead to the rephasing spectroscopic signals 
$[S_{\mathrm{FS}}]^{\omega_1,\omega_2,\omega_3,\omega_4}_{
\mathbf{e}_1,\mathbf{e}_2,\mathbf{e}_3,\mathbf{e}_4}(\tau,T,t)$  in Eq.~(\ref{equ:SFS}).
The 2D-FS signals are synchronously-phase detected with respect to the modulated laser fields at
frequency $\phi_{\mathrm{rep}} = -\phi_1 + \phi_2 + \phi_3 - \phi_4$.
In the lower and upper panels, the diagrams are grouped according to the probability 
they occur $C_{\omega_1}^p C_{\omega_2}^q$ and $C_{\omega_3}^p C_{\omega_4}^q$,
respectively. }
\label{fig:Feynman2DFS}
\end{figure*}
\end{center}

The contributions to $\hat{\rho}^{\omega_1,\omega_2}_{\mathbf{e_1},\mathbf{e_2}}(0)$ in 
Eq.~(\ref{equ:InitialState}) are clearly depicted in the lower panel of Fig.~\ref{fig:Feynman2DFS}
(see Figs.~A, B, C and D).
Starting from the system the ground state $\prjctr{g}{g}$, in the RWA and selecting only those 
contributions with $-\phi_1 + \phi_2$, the first pulse can only excite the bra and then creates an 
the optical coherence $\prjctr{g}{p}$ with probability $C^p_{\omega_1}$.
This coherence $\prjctr{g}{p}$ evolves under the action of $\mathcal{G}_{gp}(\tau)$ for a time $\tau$
when the second pulse  prepares the state $\prjctr{q}{p}$ with probability $C^q_{\omega_2}$
or a hole $-\prjctr{g}{g}$ with probability $C^p_{\omega_2}$.

As in the case of QPT based on 2D-PES \cite{YK&11,YA11}, to prepare the set
of four linearly independent states in Eq.~(\ref{equ:InitialState})  (see also Figs.~1.A--1.D), it
suffices to consider a pulse toolbox of two waveforms with carrier frequencies $\{\omega_+,\omega_-\}$
that create $\ket{e}$ and $\ket{e'}$ with different amplitudes.
Of course, the discrimination in the preparation of $\ket{e}$ and  $\ket{e^\prime}$ depends on how
close to resonance the carrier frequencies are.
For an extensive and detailed analysis on this respect, see Refs.~\cite{YK&11,YA11}.

\textit{Evolution}\textemdash
Once the initial state $\hat{\rho}^{\omega_1,\omega_2}_{\mathbf{e_1},\mathbf{e_2}}(0)$ is effectively 
prepared, i.e., after the action of the first-two pulses $T\gtrsim3\sigma$, the system evolves over a 
time $T$ under the action of the super operator $\hat{\chi}(T)$, 
according to
\begin{equation}
\label{equ:rhoT}
\hat{\rho}^{\omega_1,\omega_2}_{\mathbf{e_1},\mathbf{e_2}}(T) = 
\hat{\chi}(T) \hat{\rho}^{\omega_1,\omega_2}_{\mathbf{e_1},\mathbf{e_2}}(0).
\end{equation}
To avoid contamination of the initial state by terms proportional to a hole every time there is a 
single-exciton population $\prjctr{p}{p}$, it is assumed that 
$\langle ab| \hat{\chi}(T) | gg \rangle = \chi_{abgg}(T) = \delta_{ag}\delta_{bg}$, which is equivalent
to neglect processes where phonons induce upward optical transitions and spontaneous excitation 
from the single to the double exciton manifolds \cite{YK&11}.
Up to this condition, $\hat{\chi}(T)$ in Eq.~(\ref{equ:rhoT}) describes the dynamics induced by any 
bath model and accounts for any system-bath coupling.

\textit{Final State Detection}\textemdash
The very nature of the fluorescence detection in 2D-FS suggests considering contributions
from excitation configurations that lead to populations only. 
In the upper panel of Fig.~\ref{fig:Feynman2DFS}, those contributions are schematically displayed
and grouped according to the probability $C_{\omega_3}^p C_{\omega_4}^q$ that they occur.

In contrast to QPT, which is based on 2D-PES, there are here fourteen possibilities for the final 
state instead of ten. 
Thus, twenty independent experiments are needed instead of sixteen.
This comes at the expense of the different role that the fourth pulse has in each technique, namely, 
heterodyne detection in 2D-PES and the generation of populations in 2D-FS.
However, the signals that lead to population of the doubly-excited exciton manifold 
$\left| \makebox[11pt]{$f$} \rangle \langle \makebox[11pt]{$f$} \right|$ from the coherence
$\left| \makebox[11pt]{$f$} \rangle \langle \makebox[11pt]{$e$} \right|$ must be summed up
to the signal that lead to the population 
$\left| \makebox[11pt]{$e$} \rangle \langle \makebox[11pt]{$e$} \right|$.
In the summation, the process $\left| \makebox[11pt]{$f$} \rangle \langle \makebox[11pt]{$e$} \right|
\rightarrow \left| \makebox[11pt]{$f$} \rangle \langle \makebox[11pt]{$f$} \right|$ is weighted with a 
factor $\Gamma$, while the process $\left| \makebox[11pt]{$f$} \rangle \langle \makebox[11pt]{$e$} \right|
\rightarrow \left| \makebox[11pt]{$e$} \rangle \langle \makebox[11pt]{$e$} \right|$ has weight -1.
The same procedure applies to the process $\left| \makebox[11pt]{$f$} \rangle \langle \makebox[11pt]{$e^\prime$} \right|
\rightarrow \left| \makebox[11pt]{$f$} \rangle \langle \makebox[11pt]{$f$} \right|$ and 
$\left| \makebox[11pt]{$f$} \rangle \langle \makebox[11pt]{$e^\prime$} \right|
\rightarrow \left| \makebox[11pt]{$e^\prime$} \rangle \langle \makebox[11pt]{$e^\prime$} \right|$.
This procedure leads to the sixteen independent signals that are needed to reconstruct the sixteen 
elements of the process tensor $\chi_{nm\nu\mu}(T)$.

By using the same toolbox as in the preparation state and following closely the notation in 
Ref.~\cite{YK&11}, the total signal is given by
\begin{equation}
\label{equ:SFS}
\begin{split}
[S_{\mathrm{FS}}]&^{\omega_1,\omega_2,\omega_3,\omega_4}_{
\mathbf{e}_1,\mathbf{e}_2,\mathbf{e}_3,\mathbf{e}_4}(\tau,T,t) 
\\
= 
&\sum_{p,q,r,s\in\{e,e'\}} \hspace{-0.25cm} C^p_{\omega_1} C^q_{\omega_2} C^r_{\omega_3} 
C^s_{\omega_4}
P^{p,q,r,s}_{\mathbf{e}_1,\mathbf{e}_2,\mathbf{e}_3,\mathbf{e}_4}
\end{split}
\end{equation}
with
\begin{equation}
\label{equ:Ppqee}
\begin{split}
&\hspace{-0.75cm}P^{p,q,e,e}_{\mathbf{e}_1,\mathbf{e}_2,\mathbf{e}_3,\mathbf{e}_4}(\tau,T,t)=
\\
&\left(\boldsymbol{\mu}_{pg} \cdot \mathbf{e}_1\right) \left(\boldsymbol{\mu}_{qg} \cdot \mathbf{e}_2\right)
\mathscr{G}_{gp}(\tau)
\\ &\times 
\left\{ 
\left(\boldsymbol{\mu}_{e g} \cdot \mathbf{e}_3\right) 
\left(\boldsymbol{\mu}_{e g} \cdot \mathbf{e}_4\right) \mathscr{G}_{e g}(t)
\right.
\\
&\left. \times \left[\chi_{qqqp}(T) -\delta_{pq} - \chi_{e e qp}(T) \right] -(1-\Gamma)
\right. 
\\
&\times \left. \left(\boldsymbol{\mu}_{f e'} \cdot \mathbf{e}_3\right) 
\left(\boldsymbol{\mu}_{f e'} \cdot \mathbf{e}_4\right) 
\mathscr{G}_{e g}(t) \chi_{e' e' qp}(T)
\right\}
\end{split}
\end{equation}
and
\begin{equation}
\label{equ:Ppqeep}
\begin{split}
&\hspace{-1.25cm}P^{p,q,e,e'}_{\mathbf{e}_1,\mathbf{e}_2,\mathbf{e}_3,\mathbf{e}_4}(\tau,T,t)=
\\
&-\left(\boldsymbol{\mu}_{pg} \cdot \mathbf{e}_1\right) \left(\boldsymbol{\mu}_{qg} \cdot \mathbf{e}_2\right)
\mathscr{G}_{gp}(\tau)
\\ 
&\times 
\left\{ 
\left(\boldsymbol{\mu}_{e g} \cdot \mathbf{e}_3\right) 
\left(\boldsymbol{\mu}_{e' g} \cdot \mathbf{e}_4\right) 
\mathscr{G}_{e' g}(t) \chi_{e' e qp}(T)
\right.
\\
&+ (1-\Gamma) \left. \left(\boldsymbol{\mu}_{f e'} \cdot \mathbf{e}_3\right) 
\left(\boldsymbol{\mu}_{f e} \cdot \mathbf{e}_4\right) \mathscr{G}_{f e}(t)
\right\}
\end{split}
\end{equation}
Analogous expressions hold for $P^{p,q,e',e'}_{\mathbf{e}_1,\mathbf{e}_2,\mathbf{e}_3,\mathbf{e}_4}$ 
and $P^{p,q,e',e}_{\mathbf{e}_1,\mathbf{e}_2,\mathbf{e}_3,\mathbf{e}_4}$ after interchanging 
$e \leftrightarrow e'$.
The 2D-FS signals $[S_{\mathrm{FS}}]^{\omega_1,\omega_2,\omega_3,\omega_4}_{
\mathbf{e}_1,\mathbf{e}_2,\mathbf{e}_3,\mathbf{e}_4}(\tau,T,t)$ in Eq.~(\ref{equ:SFS}) with (\ref{equ:Ppqee}) 
and (\ref{equ:Ppqeep}) are the main result of this article.
They allow for the reconstruction of the dynamics of excitonic systems based on 2D-FS that
is an attractive approach to reach QPT at the level of single molecules.
Remarkably, the appealing form of Eqs.~(\ref{equ:SFS}), (\ref{equ:Ppqee}) 
and (\ref{equ:Ppqeep}) allows for an immediate connection with the protocol derived in
Refs.~\cite{YK&11,YA11}.
Specifically, up to a global minus sign that is consistent with previous investigations \cite{LP&11,PW&12}, 
results in Refs.~\cite{YK&11,YA11} are obtained by simply setting $\Gamma=2$ in Eqs.~(\ref{equ:SFS}), 
(\ref{equ:Ppqee}) and (\ref{equ:Ppqeep}). 

Because the probed sample is an ensemble of isotropically distributed molecules in solution, 
an isotropic average of 
$\left(\boldsymbol{\mu}_{a} \cdot \mathbf{e}_1\right) \left(\boldsymbol{\mu}_{b} \cdot \mathbf{e}_2\right)
\left(\boldsymbol{\mu}_{c} \cdot \mathbf{e}_3\right) \left(\boldsymbol{\mu}_{d} \cdot \mathbf{e}_4\right)$
is needed. 
In doing so, standard procedures are followed (see, e.g., Chap.~11 in Ref.~\cite{CT12} or Sec.~3.3 
in Ref.~\citenum{Cho10}). 
Specifically, the isotropic average is given by
\begin{equation}
\begin{split}
&\langle \left(\boldsymbol{\mu}_{a} \cdot \mathbf{e}_1\right) \left(\boldsymbol{\mu}_{b} \cdot \mathbf{e}_2\right)
\left(\boldsymbol{\mu}_{c} \cdot \mathbf{e}_3\right) \left(\boldsymbol{\mu}_{d} \cdot \mathbf{e}_4\right) \rangle_{\mathrm{iso}}
\\
&= \sum_{m_1,m_2,m_3,m_4} \mathsf{I}^{(4)}_{e_1,e_2,e_3,e_4,m_1,m_2,m_3,m_4}
\\
&\times
\left(\boldsymbol{\mu}_{a} \cdot \mathbf{m}_1\right) \left(\boldsymbol{\mu}_{b} \cdot \mathbf{m}_2\right)
\left(\boldsymbol{\mu}_{c} \cdot \mathbf{m}_3\right) \left(\boldsymbol{\mu}_{d} \cdot \mathbf{m}_4\right).
\end{split}
\end{equation}
where $\mathbf{e}_i$ and $\mathbf{m}_i$ denote the polarization of the pulses in the laboratory and 
molecule-fixed frames, respectively. $e_i=\{e_{xi},e_{yi},e_{xi}\}$ and $m_i=\{m_{xi},m_{yi},m_{xi}\}$ 
are the components of the polarization vectors  $\mathbf{e}_i$ and $\mathbf{m}_i$, respectively.
The isotropically invariant tensor $\mathsf{I}^{(4)}$ is given by
\begin{equation}
\begin{split}
&\hspace{-1.25cm}\mathsf{I}^{(4)}_{e_1,e_2,e_3,e_4, m_1,m_2,m_3,m_4} =
\\
&\frac{1}{30} \left( \delta_{e_1e_2} \delta_{e_3e_4}\quad  \delta_{e_1e_3} \delta_{e_2e_4} \quad
 \delta_{e_1e_4} \delta_{e_2e_3} \right)
\\
& \times 
\left( \begin{array}{ccc}
4 & -1 & -1 \\
-1 & 4 & -1 \\
-1 & -1 & 4 \end{array} \right)
\left( \begin{array}{c}
 \delta_{m_1m_2} \delta_{m_3m_4} \\
\delta_{m_1m_3} \delta_{m_2m_4}\\
\delta_{m_1m_4} \delta_{m_2m_3}  \end{array} \right).
\end{split}
\end{equation}
The explicit expression for the relevant case of interest in the collinear configuration 
used in 2D-FS, $\mathbf{e_1}=\mathbf{e_2}=\mathbf{e_3}=\mathbf{e_4}=\mathbf{z}$,
can be found in Appendix~\ref{app:IsoAverg}.
Thus, after isotropically averaging,
\begin{equation}
\label{equ:avergSFS}
\begin{split}
\langle [S_{\mathrm{FS}}]&^{\omega_1,\omega_2,\omega_3,\omega_4}_{
\mathbf{e}_1,\mathbf{e}_2,\mathbf{e}_3,\mathbf{e}_4}(\tau,T,t) \rangle_{\mathrm{iso}}
\\
= 
&\sum_{p,q,r,s\in\{e,e'\}} \hspace{-0.25cm} C^p_{\omega_1} C^q_{\omega_2} C^r_{\omega_3} 
C^s_{\omega_4}
\langle P^{p,q,r,s}_{\mathbf{e}_1,\mathbf{e}_2,\mathbf{e}_3,\mathbf{e}_4}\rangle_{\mathrm{iso}}.
\end{split}
\end{equation}
Because in 2D-FS the laser pulses are collinear, it is possible to set $\mathbf{e}_j = \mathbf{z}$
at this point.
Additionally, it is assumed below that $\tau=0$ and $t=0$ so that only the signals 
$P^{p,q,r,s}_{\mathbf{e}_1,\mathbf{e}_2,\mathbf{e}_3,\mathbf{e}_4}(0,T,0)$ are considered.
However, following a similar procedure as in Ref.~\cite{YK&11}, this restriction can be relaxed 
and arbitrary $\tau$ and $t$ can be considered.
For the sake of generality, the polarization vectors are denoted independently by $\mathbf{e}_j$ 
and $\tau$ and $t$ are not set to zero in the above expressions.

The extraction procedure of the matrix elements of $\hat{\chi}$ from 
$\langle [S_{\mathrm{FS}}]^{\omega_1,\omega_2,\omega_3,\omega_4}_{
\mathbf{e}_1,\mathbf{e}_2,\mathbf{e}_3,\mathbf{e}_4}(\tau,T,t) \rangle_{\mathrm{iso}}$ follows 
from Eqs.~(\ref{equ:avergSFS}), (\ref{equ:Ppqee}) and (\ref{equ:Ppqeep}). 
Note that in doing so the sixteen 2D-FS signals $\langle [S_{\mathrm{FS}}]^{\omega_1,\omega_2,\omega_3,\omega_4}_{
\mathbf{e}_1,\mathbf{e}_2,\mathbf{e}_3,\mathbf{e}_4}(\tau,T,t) \rangle_{\mathrm{iso}}$, and the 
sixteen auxiliary signals $\langle P^{p,q,r,s}_{\mathbf{e}_1,\mathbf{e}_2,\mathbf{e}_3,\mathbf{e}_4}(\tau,T,t)
\rangle_{\mathrm{iso}}$
can be grouped into the sixteen-dimensional vectors $\langle [\mathsf{S}_{\mathrm{FS}}] (\tau,T,t) \rangle_{\mathrm{iso}}$
and $\langle \mathsf{P}(\tau,T,t) \rangle_{\mathrm{iso}}$, respectively.
This allows writing Eq.~(\ref{equ:avergSFS}) as 
$\langle [\mathsf{S}_{\mathrm{FS}}] (\tau,T,t) \rangle_{\mathrm{iso}} = \mathsf{C}
\langle \mathsf{P}(\tau,T,t)\rangle_{\mathrm{iso}}$, where the matrix elements of $\mathsf{C}$ contains 
the probabilities $C^p_{\omega_1} C^q_{\omega_2} C^r_{\omega_3} C^s_{\omega_4}$.
Then, the first step in the extraction procedure is to invert the matrix $\mathsf{C}$ so that the
signals $\langle P^{p,q,r,s}_{\mathbf{e}_1,\mathbf{e}_2,\mathbf{e}_3,\mathbf{e}_4}\rangle_{\mathrm{iso}}$
can be extracted from the measured signals $\langle [S_{\mathrm{FS}}]^{\omega_1,\omega_2,\omega_3,\omega_4}_{
\mathbf{e}_1,\mathbf{e}_2,\mathbf{e}_3,\mathbf{e}_4}(\tau,T,t) \rangle_{\mathrm{iso}}$, i.e.,
$\langle \mathsf{P}(\tau,T,t)\rangle_{\mathrm{iso}} = \mathsf{C}^{-1} 
\langle [\mathsf{S}_{\mathrm{FS}}] (\tau,T,t) \rangle_{\mathrm{iso}} $.
The second step comprises the extraction of the sixteen elements of the process tensor $\hat{\chi}(T)$
from the isotropically-averaged version of Eqs.~(\ref{equ:Ppqee}) and (\ref{equ:Ppqeep}).
This process can be accomplished by conveniently defining a sixteen-dimensional vector
$\boldsymbol{\chi}(T)$, such that $\boldsymbol{\chi}(T) = 
\mathsf{M}^{-1} \langle \mathsf{P}(0,T,0)\rangle_{\mathrm{iso}}$.
See the Appendices for further details.

\section{Numerical Example}
As a concrete example, consider parameters of  relevance in the context of light-harvesting systems 
\cite{KBP06,LCF07}.
Specifically, to compare with previous results \cite{YK&11}, consider $\varpi_1 = 12881$~cm$^{-1}$, 
$\varpi_2 = 12719$~cm$^{-1}$, $J=120$~cm$^{-1}$,  $d_2/d_1=2$ and $\phi=0.3$.
The two-waveform toolbox is assumed to have frequencies $\omega_+ = 13480$~cm$^{-1}$ and 
$\omega_- = 12130$~cm$^{-1}$ so that $\varpi_i=\{\omega_+,\omega_-\},\,\forall\,i$ and pulse
width $\sigma=40$~fs.
To simulate the signals, the spectral density $J_n(\omega)$ of the local vibrational environments 
are assumed identical and given by 
$J_n(\omega) = (\lambda_{\mathrm{r.e.}}/\omega_{\mathrm{c}})\omega \exp(-\omega/\omega_{\mathrm{c}})$ 
where the cutoff frequency is set as $\omega_{\mathrm{c}} = 120$~cm$^{-1}$ while the reorganization 
energy is chosen as $\lambda_{\mathrm{r.e.}} = 30$~cm$^{-1}$.
These set of parameters are relevant for light-harvesting systems and were used in Ref.~\cite{YK&11}. 

In the simulations below, an inhomogeneously broadened ensemble of $10^4$ dimers with diagonal
disorder is considered. 
Specifically, it is assumed that the site energies $\varpi_1^\prime$ and $\varpi_2^\prime$ in the 
ensemble follow a Gaussian distribution centered at $\varpi_1$ and $\varpi_2$  with standard deviation 
$\sigma_{\mathrm{inh}}=40$~cm$^{-1}$. 
The dynamics are solved at the level of the secular Redfield master equation at room temperature
and for $T\ge 3\sigma$.

\begin{figure}[h]
\includegraphics[width=\columnwidth]{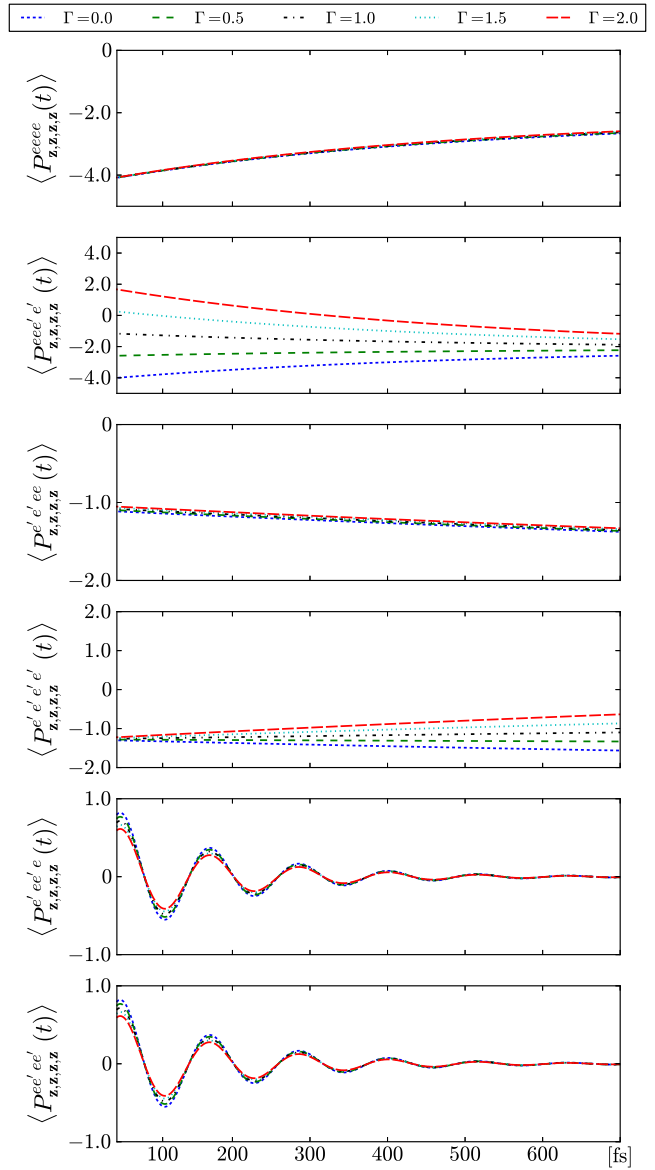}
\caption{Nonvanishing $\langle P^{p,q,r,s}_{\mathbf{z},\mathbf{z},\mathbf{z},\mathbf{z}}(0,T,0)\rangle_{\mathrm{iso}}$
signals for a variety of values of $\Gamma$ and for $3\sigma \le T \le 700$~fs.
Curves with $\Gamma=2$ coincides with those extracted in Ref.~\cite{YK&11}.}
\label{fig:Ppqrs}
\end{figure}
Figure~\ref{fig:Ppqrs} depicts the nonvanishing real parts of 
$\langle P^{p,q,r,s}_{\mathbf{z},\mathbf{z},\mathbf{z},\mathbf{z}}\rangle_{\mathrm{iso}}$ for a variety
of values of the quantum yields parameter $\Gamma$.
From the functional dependence on $\Gamma$ of the signals 
$\langle P^{p,q,r,s}_{\mathbf{z},\mathbf{z},\mathbf{z},\mathbf{z}}\rangle_{\mathrm{iso}}$ 
in Eqs.~(\ref{equ:Ppqee}) and (\ref{equ:Ppqeep}), three cases are of interest: 
(i) For $\Gamma=0$, the contribution from the excited state absorption (ESA) 
pathways has the same sign as the stimulated emission (SE) and  ground-state bleach (GSB) contributions 
(see the double-sided Feynman's diagrams in Fig.~\ref{fig:Feynman2DFS} or the discussion in Ref.~\cite{PW&12}). 
Thus, the amplitude of the signals is the largest possible.
(ii) For $\Gamma=1$, the ESA pathways do not contribute to the signal and the amplitude of the 
signals is expected to be smaller than in the case of $\Gamma=0$.
(iii) For $\Gamma=2$,  the contribution from the ESA pathways has opposite sign to the SE and  GSB 
contributions, so that the amplitude is expected to be smaller than in the previous case  $\Gamma=1$ .
These expectations are confirmed by simulations in Fig.~\ref{fig:Ppqrs}.
For completeness, the intermediate cases $\Gamma=0.5$ and $\Gamma=1.5$ were also 
depicted in Fig.~\ref{fig:Ppqrs}.

\begin{figure}[h]
\includegraphics[width=\columnwidth]{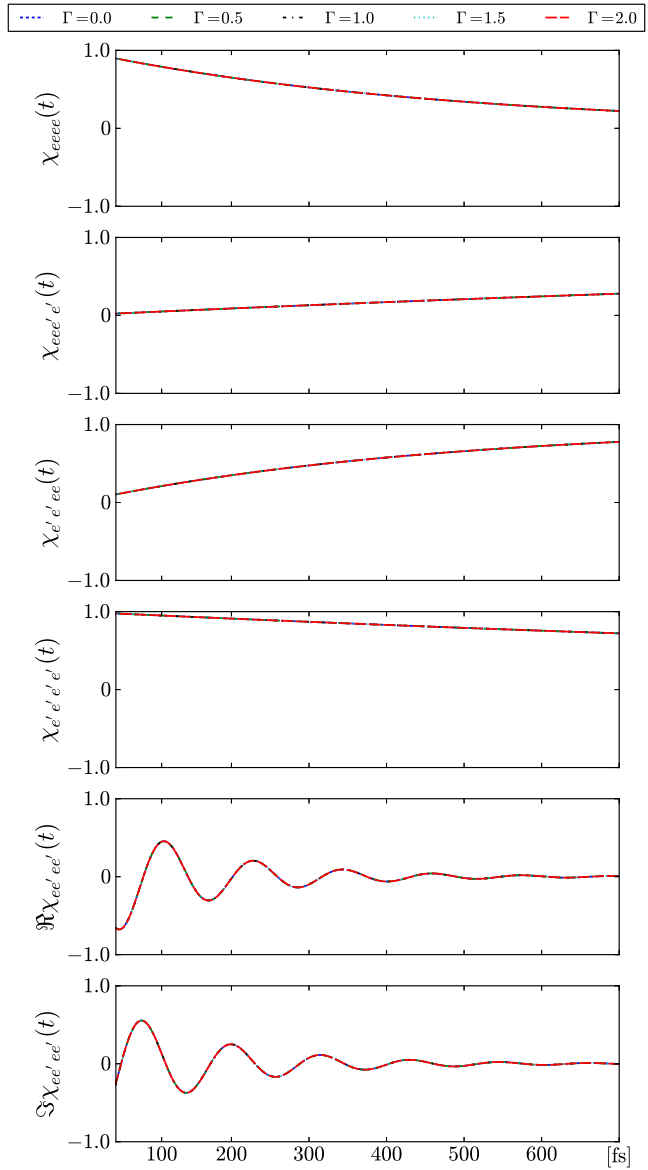}
\caption{Nonvanishing elements of the QPT tensor $\chi_{nm\nu\mu}(T)$ obtained form the
$\langle P^{p,q,r,s}_{\mathbf{z},\mathbf{z},\mathbf{z},\mathbf{z}}\rangle_{\mathrm{iso}}$ depicted
in Fig.~\ref{fig:Ppqrs} for a variety of values of $\Gamma$ and for $3\sigma \le T \le 700$~fs.
Curves with $\Gamma=2$ coincides with those extracted in Ref.~\cite{YK&11}.}
\label{fig:Xpqrs}
\end{figure}

Based on the signals $\langle P^{p,q,r,s}_{\mathbf{z},\mathbf{z},\mathbf{z},\mathbf{z}}\rangle_{\mathrm{iso}}$ 
obtained above, the QPT tensor is reconstructed in Fig.~\ref{fig:Xpqrs}.
The reconstruction appears to be insensitive to the value of the quantum yield parameter $\Gamma$.
This unexpected result can be understood after noticing that the QPT tensor $\chi_{nm\nu\mu}(T)$ is a 
characteristic of the singly exited exciton manifold and $\Gamma$ is a function of the doubly-excited 
exciton manifold.
Thus, QPT of the singly-exited exciton manifold by 2D-FS is robust against nonradiative processes
of the doubly-excited exciton manifold and benefits from the quality of the signals discussed above.

Results in Fig.~\ref{fig:Xpqrs} agree with the secular Redfield tensor used to simulate the 2D-FS 
signals.
If the signal were obtained from experimental data, a careful analysis of the propagation of errors 
would be in order. 
In particular, it is necessary to include fluctuations in the laser intensity at each time $T$ at which
the signals are collected, and to pay attention to the stability conditions imposed by the invertibility 
of the matrices $\mathsf{C}$ and $\mathsf{M}$ \cite{YK&11}.  
In this article, interest was in providing a proof-of-principle for the scheme derived above, so that 
Fig.~\ref{fig:Xpqrs} is aimed to depict the type of information that can be extracted from the protocol.

Specifically, (i) if for a particular photochemical system, non-negligible, non-secular terms emerge 
during the reconstruction of the process tensor $\hat{\chi}$ from experimental data, that would imply, 
e.g, that coherent control schemes assisted by the environment \cite{PYB13,PB13b} may be 
applied in that particular system.
(ii) If the decay of the tensor elements associated to the coherences of the density matrix,
$\chi_{nmnm}$ with $n\ne m$, are non-exponential, it may indicate the presence of non-Markovian
dynamics \cite{PB14b}.
The deviation from exponencial decay behavior may even be considered as a measure of 
the non-Markovian character of the dynamics-- a relevant topic in the context of open quantum 
systems.
(iii) Although in multilevel systems the decay rate of the elements $\chi_{nmnm}$ cannot be 
directly associated to the decay rate of the coherences $\langle n | \hat{\rho}| m \rangle$ of the 
density matrix, the decay rate of $\chi_{nmnm}$ provides information about the lifetime of particular 
transfer and coherent mechanisms.

\section{Discussion}
%
Having experimental access to the process tomography tensor $\chi_{nm\nu\mu}(t)$ is fundamental
to revealing energy pathways in exciton dynamics, and in designing control strategies to increase 
transport efficiency.
Specifically, applications of QPT to photosynthetic light-harvesting systems can
(i) rule out certain transfer mechanisms proposed in the literature, and 
(ii) address the question about the quantum/classical nature of the energy transport in certain biological 
systems from an experimental viewpoint. 
In addressing these issues, a complete analysis of the classical/quantum correlations encoded in the 
process tomography tensor, as well as an analysis of the main contributing elements to energy transport
is required and will be discussed elsewhere.

A variety of applied and foundational problems can be addressed once the process tomography tensor
is reconstructed.
From a foundational viewpoint, if the process tomography tensor is translated into the phase-space
representation of quantum mechanics, it reduces to the propagator of the Wigner function \cite{DP09,DGP10,PID10}. 
Based on this object, it is possible to \emph{experimentally} reconstruct signatures of quantum chaos 
such as scars with \emph{sub-Planckian resolution} \cite{DP09}.
Phase-space resolution below $\hbar$ can be achieved here because the process tomography tensor,
or equivalently the propagator of the Wigner function, is not a physical state and therefore, it is not restricted 
by the uncertainty principle \cite{DP09}.

In the same way that 2D-PES was extended to study chemical exchange
to obtain reaction rates under well controlled conditions (see, e.g., Chap.~10 in Ref.~\citenum{Cho10}), a 
straightforward extension of QPT is the accurate measurement of concentration of different species in 
chemical reactions.
This has been considered very recently in the literature \cite{CL&12}.
In this context, interest is in the population dynamics $\chi_{nn\nu\nu}(T)$ of the different chemical species,
which under Markovian dynamics are in accordance to detailed balance and Onsager's regression hypothesis
\cite{PT&14,Tal86,FO96}.
In this respect, because ultrafast spectroscopy allows for the study of chemical exchange with no need of 
pressure, temperature, pH nor concentration jumps, it is expected that the proposed approach provides 
experimental evidence for the failure of the Onsager's regression hypothesis induced by non-Markovian 
dynamics at the quantum level \cite{PT&14,Tal86,FO96}.

The QPT scheme introduced here can be readily implemented at the experimental level, and constitutes 
a first step toward the formulation of QPT at the single-molecule level.
Such a scheme would certainly incorporate quantum aspects of the electromagnetic radiation such as the 
use of time-energy-entangled photons \cite{RM&13}. This is already under development in our laboratories. 
Finally, based on present non-linear optical activity spectroscopy (see, e.g., Chap.~16 in Ref.~\citenum{Cho10}),
by introducing time-polarization-entangled photons, instead of time-energy-entangled photons \cite{RM&13}, 
single-molecule QPT may be extended to study optically active materials at the single-molecule level.
These materials exhibit unique optical properties, and are constantly finding applications in science and industry.

%
%
%

\begin{acknowledgements}
Discussions with  Keith Nelson and Joel Yuen-Zhou are acknowledged with pleasure.
This work was supported by the Center for Excitonics, an Energy Frontier Research Center funded 
by the U.S. Department of Energy, Office of Science and Office of Basic Energy Sciences, under Award 
Number DE-SC0001088, by \textit{Comit\'e para el Desarrollo de la Investigaci\'on}
--CODI-- of Universidad de Antioquia, Colombia under the \textit{Estrategia de Sostenibilidad} 2015-2016, 
by the \textit{Colombian Institute for the Science and Technology Development} --COLCIENCIAS--
under the contract number 111556934912 and by the National Science Foundation, Chemistry of Life 
Processes Program (CHE-1307272- to A.H.M).
\end{acknowledgements}

\bibliography{qptv2}

\begin{widetext}
\appendix

\section{Initial State Preparation}
\label{app:InitalPrep}
Because the effective initial state $\hat{\rho}_{\mathbf{e}_1,\mathbf{e}_1}^{\omega_1,\omega_2}(t+T)$
is prepared by the first two pulses, it is of second order in $\lambda$.
Thus, after applying second order perturbation theory to the time evolution of the system density matrix
(see, e.g., Chap.~5 in Ref.~\cite{Muk99}), the effective initial state reads
\begin{equation}
\hat{\rho}_{\mathbf{e}_1,\mathbf{e}_1}^{\omega_1,\omega_2}(t+T) =
\left(\frac{1}{\mathrm{i}} \right)^2 
\int_{-\infty}^{t_2+T} \mathrm{d}t'' \int_{-\infty}^{t''} \mathrm{d}t' 
\mathscr{G}_2(t_2+T,t'') \mathscr{V}(t'') \mathscr{G}_1(t'',t') \mathscr{V}(t')
| g \rangle \langle g |,
\end{equation}
where it was assumed that $\hat{\rho}(-\infty) = | g \rangle \langle g |$. 
Symbols in calligraphic font denote superoperators; in particular, 
$
\tilde{\mathscr{V}} = \sum_{i=1}^4 \hat{\mathscr{V}}_i(t),
$
where $\tilde{\mathscr{V}}_i = [\hat{V}_i,\cdot]$.
Assuming that the rotating wave approximation (RWA) holds, and that the rephasing 
signal is synchronously-phase detected at $\phi = -\phi_1 + \phi_2 + \phi_3 -\phi_4$, 
the interaction with the electromagnetic radiation is conveniently described by
\begin{align}
\hat{V}_1 &= -\lambda \hat{\boldsymbol{\mu}}^< \cdot \mathbf{e}_1 E(t-t_1) 
\mathrm{e}^{\mathrm{i} \omega_1 (t-t_1)},
\\ 
\hat{V}_2 &= -\lambda \hat{\boldsymbol{\mu}}^> \cdot \mathbf{e}_2 E(t-t_2) 
\mathrm{e}^{-\mathrm{i} \omega_2 (t-t_2)},
\\
\hat{V}_3 &= -\lambda \hat{\boldsymbol{\mu}}^> \cdot \mathbf{e}_3 E(t-t_3) 
\mathrm{e}^{-\mathrm{i} \omega_3 (t-t_3)},
\\
\hat{V}_4 &= -\lambda \hat{\boldsymbol{\mu}}^< \cdot \mathbf{e}_4 E(t-t_4) 
\mathrm{e}^{\mathrm{i} \omega_4 (t-t_4)},
\end{align}
where $\hat{\boldsymbol{\mu}}^< = \sum_{\varpi_p<\varpi_q} 
\boldsymbol{\mu}_{pq} |p\rangle \langle q|$ promotes emissions from the ket and
absorptions on the bra, and $ \hat{\boldsymbol{\mu}}^> = (\hat{\boldsymbol{\mu}}^{<})^{\dag}$
induces the opposite processes.

In the following, it is considered that the first and the second pulse are well separated, i.e., 
$\tau = t_2 -t_1 > 3 \sigma$.
This allows for the substitutions $\mathscr{V}(t'') =  \mathscr{V}_2(t'')$ and 
$\mathscr{V}(t'') =  \mathscr{V}_1(t')$.

\begin{equation}
\begin{split}
\hat{\rho}_{\mathbf{e}_1,\mathbf{e}_1}^{\omega_1,\omega_2}(t+T) = 
\lambda^2 \sum_{pq} 
& 
\left\{
	\int\limits_{-\infty}^{t_2+T} \mathrm{d}t'' \chi(T) 
	\left[ 
		\mathscr{G}_{qp}(t_2 - t'') 
		\left( 
			\boldsymbol{\mu}_{qg} \cdot \mathbf{e}_2 \prjctr{q}{g} 
			E(t'' - t_2) \eu^{\iu \omega_1 (t'' - t_2)}
		\right) \phantom{\int\limits_{-\infty}^{t''} \mathrm{d}t' }
	\right. 
\right.
	\\
& 
\left. \left. 
		\times 
		\int\limits_{-\infty}^{t''} \mathrm{d}t'  \mathscr{G}_{gp}(t'' - t') \prjctr{g}{g}
		\left( 
			\boldsymbol{\mu}_{qg} \cdot \mathbf{e}_2 \prjctr{q}{g} 
			E(t'' - t_2) \eu^{-\iu \omega_2 (t'' - t_2)}
		\right) 
	\right] 
\right.
\\
&\hspace{-1cm}-  
\left.  
	\int\limits_{-\infty}^{t_2+T}  \mathrm{d}t'' \chi(T) 
		\left[
			\mathscr{G}_{gg}(t_2 - t'') 
			 \int\limits_{-\infty}^{t''} \mathrm{d}t' \mathscr{G}_{gp} (t'' - t') 
			\prjctr{g}{g}  
			\left( 
				\boldsymbol{\mu}_{pg} \cdot \mathbf{e}_1 \prjctr{q}{g} 
				E(t' - t_1) \eu^{\iu \omega_1 (t' - t_1)}
			\right)
		\right. 
\right.
\\
&
\left. 
		\left. 
		\times 
		\left( 
			\boldsymbol{\mu}_{qg} \cdot \mathbf{e_1} \prjctr{q}{g} 
			E(t'' - t_2) \eu^{-\iu \omega_2 (t'' - t_2)}
		\right) \phantom{\int\limits_{-\infty}^{t''} \mathrm{d}t' }\hspace{-1cm}
	\right] 
\right\}.
\end{split}
\end{equation}

If the duration of the pulse $\sigma$ is much sorter than the dynamics induced by the environment
characterized by $\Gamma_{nm}$, i.e., if $\sigma \ll \Gamma_{nm}^{-1}$, then decoherering contributions
can be neglected so that $\mathcal{G}_{gp}(t_1-t') \approx \exp\left[\mathrm{i} \varpi_{pg}(t_1-t') \right]$,
$\mathcal{G}_{qp}(t_2-t'')\mathcal{G}_{gp}(t''-t_2) \approx \exp\left[-\mathrm{i} \varpi_{qp}(t_2-t'')\right]
\exp\left[\mathrm{i} \varpi_{pg}(t_2-t'')\right] =  \exp\left[-\mathrm{i} \varpi_{qg}(t_2-t'')\right]$,
$\mathcal{G}_{gp}(t''-t_2) \approx \exp\left[-\mathrm{i} \varpi_{pg}(t''-t_2) \right]$ and
$\mathcal{G}_{gg}(t''-t_2) \approx \exp\left[-\mathrm{i} \varpi_{gg}(t''-t_2) \right]$.
Moreover, $\mathscr{G}_{gp} (t'' - t') \approx \mathscr{G}_{gp} (t'' - t_2) \mathscr{G}_{gp} (t_1 - t') \mathscr{G}_{gp} (t_2' - t_1)$.
After some manipulations,
\begin{equation}
\hat{\rho}_{\mathbf{e}_1,\mathbf{e}_1}^{\omega_1,\omega_2}(t+T) \approx 
-\chi(T) \sum_{pq} 
\left\{ C^p_{\omega_1} C^q_{\omega_2} (\boldsymbol{\mu}_{pg}\cdot \mathbf{e}_1)
(\boldsymbol{\mu}_{qg}\cdot \mathbf{e}_2) \mathscr{G}_{gp} (\tau)
\left(\prjctr{q}{p} - \delta_{pq}\prjctr{g}{g}\right)
\right\}
\end{equation}
with
\begin{equation}
C^{p}_{\omega_j} = -\frac{\lambda}{\mathrm{i}} \int_{-\infty}^{\infty} \mathrm{d}s \,
\exp[\mathrm{i}(\omega_j - \varpi_{pg})s]E(s)
= -\frac{\lambda}{\mathrm{i}} \sqrt{2\pi \sigma^2} \exp[-(\varpi_{pg}-\omega_j)^2]
\end{equation}
As mentioned in the main text, this effective initial state coincides with the one prepared
by 2D-PES in Ref.~\cite{YK&11}.

\section{Final State Detection}
\label{app:FinalPrep}
To derive the explicit form of the density operator after the action of the four pulses, it is assumed
that the third and the fourth pulses are well separated as well.
To account for the action of the third pulse and a subsequent period of free evolution, perturbation
theory is applied once more, so that the density operator of the system reads
\begin{equation}
\begin{split}
&\hspace{-1.25cm}\hat{\rho}_{\mathbf{e}_1,\mathbf{e}_2,\mathbf{e}_3}^{\omega_1,\omega_2,\omega_3}(\tau+T+t)
= \sum_{pq} 
\left\{
	\left[
		C^e_{\omega_3}   (\boldsymbol{\mu}_{eg}\cdot \mathbf{e}_3) \mathscr{G}_{eg} (t)
		\langle g \left | \hat{\rho}_{\mathbf{e}_1,\mathbf{e}_1}^{\omega_1,\omega_2}(t+T) \right | g \rangle 
		\phantom{C^{e^\prime}_{\omega_3}}
	\right. 
\right. 
\\
& 
\left. 
	\left.
		-
		C^e_{\omega_3}   (\boldsymbol{\mu}_{eg}\cdot \mathbf{e}_3) \mathscr{G}_{eg} (t)
		\langle e \left | \hat{\rho}_{\mathbf{e}_1,\mathbf{e}_1}^{\omega_1,\omega_2}(t+T) \right | e \rangle
		-
		C^{e^\prime}_{\omega_3}   (\boldsymbol{\mu}_{e^\prime g}\cdot \mathbf{e}_3) \mathscr{G}_{e^\prime g} (t)
		\langle e \left | \hat{\rho}_{\mathbf{e}_1,\mathbf{e}_1}^{\omega_1,\omega_2}(t+T) \right | e^\prime \rangle
	\right] 
	\prjctr{e}{g}
\right.
\\ 
+&
\left.
	\left[
		C^{e^\prime}_{\omega_3}   (\boldsymbol{\mu}_{e^\prime g}\cdot \mathbf{e}_3) \mathscr{G}_{e^\prime g} (t)
		\langle g \left | \hat{\rho}_{\mathbf{e}_1,\mathbf{e}_1}^{\omega_1,\omega_2}(t+T) \right | g \rangle 
		\phantom{C^{e^\prime}_{\omega_3}}
	\right. 
\right. 
\\
& 
\left. 
	\left.
		-
		C^{e^\prime}_{\omega_3}   (\boldsymbol{\mu}_{e^\prime g}\cdot \mathbf{e}_3) \mathscr{G}_{e^\prime g} (t)
		\langle e^\prime  \left | \hat{\rho}_{\mathbf{e}_1,\mathbf{e}_1}^{\omega_1,\omega_2}(t+T) \right | e^\prime \rangle
		-
		C^{e}_{\omega_3}   (\boldsymbol{\mu}_{e g}\cdot \mathbf{e}_3) \mathscr{G}_{e^\prime g} (t)
		\langle e^\prime \left | \hat{\rho}_{\mathbf{e}_1,\mathbf{e}_1}^{\omega_1,\omega_2}(t+T) \right | e \rangle
	\right] 
	\prjctr{e^\prime}{g}
\right.
\\
+&
\left.
	\left[
		C^{e^\prime}_{\omega_3}   (\boldsymbol{\mu}_{f e}\cdot \mathbf{e}_3) \mathscr{G}_{f e} (t)
		\langle e  \left | \hat{\rho}_{\mathbf{e}_1,\mathbf{e}_1}^{\omega_1,\omega_2}(t+T) \right | e \rangle
		+
		C^{e}_{\omega_3}   (\boldsymbol{\mu}_{f e^\prime}\cdot \mathbf{e}_3) \mathscr{G}_{f e} (t)
		\langle e^\prime \left | \hat{\rho}_{\mathbf{e}_1,\mathbf{e}_1}^{\omega_1,\omega_2}(t+T) \right | e \rangle
	\right] 
	\prjctr{f}{e}
\right.
\\
+&
\left.
	\left[
		C^{e^\prime}_{\omega_3}   (\boldsymbol{\mu}_{f e}\cdot \mathbf{e}_3) \mathscr{G}_{f e^\prime} (t)
		\langle e  \left | \hat{\rho}_{\mathbf{e}_1,\mathbf{e}_1}^{\omega_1,\omega_2}(t+T) \right | e^\prime \rangle
		+
		C^{e}_{\omega_3}   (\boldsymbol{\mu}_{f e^\prime}\cdot \mathbf{e}_3) \mathscr{G}_{f e^\prime} (t)
		\langle e^\prime \left | \hat{\rho}_{\mathbf{e}_1,\mathbf{e}_1}^{\omega_1,\omega_2}(t+T) \right | e^\prime \rangle
	\right] 
	\prjctr{f}{e^\prime}
\right\},
\end{split}
\end{equation}
where 
\begin{equation}
\langle i \left | \hat{\rho}_{\mathbf{e}_1,\mathbf{e}_1}^{\omega_1,\omega_2}(t+T) \right | j \rangle
= -\sum_{pq} C^p_{\omega_1} C^q_{\omega_2} (\boldsymbol{\mu}_{pg}\cdot \mathbf{e}_1)
 (\boldsymbol{\mu}_{qg}\cdot \mathbf{e}_2) \mathscr{G}_{gp} (\tau)
\left(\chi_{ijqp}(T) -\delta_{pq}\delta_{ij}\delta_{ig} \right).
\end{equation}
Finally, the fourth pulse prepares the system in the state,
\begin{equation}
\begin{split}
&\hspace{-1.25cm}
\hat{\rho}_{\mathbf{e}_1,\mathbf{e}_2,\mathbf{e}_3,\mathbf{e}_4}^{\omega_1,\omega_2,\omega_3,\omega_4}(\tau+T+t)
\\
&=\left(
	C^{e^\prime}_{\omega_4}   (\boldsymbol{\mu}_{ef}\cdot \mathbf{e}_4)
	\langle f  \left |  \hat{\rho}_{\mathbf{e}_1,\mathbf{e}_2,\mathbf{e}_3}^{\omega_1,\omega_2,\omega_3}(\tau+T+t) \right | e \rangle
	-
	C^{e}_{\omega_4}   (\boldsymbol{\mu}_{ge}\cdot \mathbf{e}_4)
	\langle g  \left |  \hat{\rho}_{\mathbf{e}_1,\mathbf{e}_2,\mathbf{e}_3}^{\omega_1,\omega_2,\omega_3}(\tau+T+t) \right | e \rangle 
\right) 	
\prjctr{e}{e}
\\ 
&+
\left(
	C^{e}_{\omega_4}   (\boldsymbol{\mu}_{e^\prime f}\cdot \mathbf{e}_4)
	\langle f  \left |  \hat{\rho}_{\mathbf{e}_1,\mathbf{e}_2,\mathbf{e}_3}^{\omega_1,\omega_2,\omega_3}(\tau+T+t) \right | e^\prime \rangle
	-
	C^{e}_{\omega_4}   (\boldsymbol{\mu}_{ge}\cdot \mathbf{e}_4)
	\langle e^\prime  \left |  \hat{\rho}_{\mathbf{e}_1,\mathbf{e}_2,\mathbf{e}_3}^{\omega_1,\omega_2,\omega_3}(\tau+T+t) \right | g \rangle 
\right) 	
\prjctr{e^\prime}{e^\prime}
\\ 
&-
\left(
	C^{e^\prime}_{\omega_4}   (\boldsymbol{\mu}_{ e e^\prime}\cdot \mathbf{e}_4)
	\langle f  \left |  \hat{\rho}_{\mathbf{e}_1,\mathbf{e}_2,\mathbf{e}_3}^{\omega_1,\omega_2,\omega_3}(\tau+T+t) \right | e \rangle
	+
	C^{e}_{\omega_4}   (\boldsymbol{\mu}_{e\prime f}\cdot \mathbf{e}_4)
	\langle f  \left |  \hat{\rho}_{\mathbf{e}_1,\mathbf{e}_2,\mathbf{e}_3}^{\omega_1,\omega_2,\omega_3}(\tau+T+t) \right | e^\prime \rangle 
\right) 	
\prjctr{f}{f}.
\end{split}
\end{equation}
Each contribution can be easily associated to the double-sided Feynman diagrams in 
Fig.~\ref{fig:Feynman2DFS} in the main text.

Once the state of the system is obtained, the spectroscopy signals 
$[S_{\mathrm{FS}}]^{\omega_1,\omega_2,\omega_3,\omega_4}_{
\mathbf{e}_1,\mathbf{e}_2,\mathbf{e}_3,\mathbf{e}_4}(\tau,T,t)$, synchronously detected at 
$\phi=-\phi_1+\phi_2+\phi_3 - \phi_4$, 
follow from the calculation of 
$\langle \hat{A}(\tau+T+t) \rangle = \mathrm{tr} \hat{A}
\hat{\rho}_{\mathbf{e}_1,\mathbf{e}_2,\mathbf{e}_3,\mathbf{e}_4}^{
\omega_1,\omega_2,\omega_3,\omega_4}(\tau+T+t)$ with 
$\hat{A}=\sum_{\nu=\{e,e',f\}} \Gamma_\nu \prjctr{\nu}{\nu}$. 
Specifically,
\begin{equation}
\label{equ:Aoft}
\begin{split}
\langle \hat{A}(\tau+T+t) \rangle = 
&-C^{e}_{\omega_4} (\boldsymbol{\mu}_{ g e}\cdot \mathbf{e}_4)
\langle e  \left |  \hat{\rho}_{\mathbf{e}_1,\mathbf{e}_2,\mathbf{e}_3}^{\omega_1,\omega_2,\omega_3}(\tau+T+t) \right | g \rangle
\\
&-C^{e^\prime}_{\omega_4} (\boldsymbol{\mu}_{ g e^\prime}\cdot \mathbf{e}_4)
\langle e^\prime  \left |  \hat{\rho}_{\mathbf{e}_1,\mathbf{e}_2,\mathbf{e}_3}^{\omega_1,\omega_2,\omega_3}(\tau+T+t) \right | g \rangle
\\
&+C^{e^\prime}_{\omega_4} (\boldsymbol{\mu}_{e f}\cdot \mathbf{e}_4)(1-\Gamma)
\langle f  \left |  \hat{\rho}_{\mathbf{e}_1,\mathbf{e}_2,\mathbf{e}_3}^{\omega_1,\omega_2,\omega_3}(\tau+T+t) \right | e \rangle
\\
&+C^{e}_{\omega_4} (\boldsymbol{\mu}_{ f e^\prime}\cdot \mathbf{e}_4)(1-\Gamma)
\langle f  \left |  \hat{\rho}_{\mathbf{e}_1,\mathbf{e}_2,\mathbf{e}_3}^{\omega_1,\omega_2,\omega_3}(\tau+T+t) \right | e^\prime \rangle.
\end{split}
\end{equation}
After replacing the explicit functional form of the density matrix elements 
$\langle \nu  \left |  \hat{\rho}_{\mathbf{e}_1,\mathbf{e}_2,\mathbf{e}_3}^{\omega_1,\omega_2,\omega_3}(\tau+T+t) \right | \nu \rangle$
in Eq.~(\ref{equ:Aoft}) and after conveniently collecting terms, Eq.~(\ref{equ:Aoft}) leads 
the 2D-FS signals in Eq.~(\ref{equ:SFS}) that are the main result of this article. 

\section{Isotropic Averages}
\label{app:IsoAverg}
Before proceeding to the calculation of the isotropic average, it is necessary to express the
dipole transition operators in the molecular frame.
In doing so, take as reference the transition dipole operator 
$\boldsymbol{\mu}_{eg} = \mu_{eg} \mathbf{m}_z$.
Hence,  
$\boldsymbol{\mu}_{e^\prime g} = \mu_{e^\prime g} \cos( \theta_{e^\prime g}) \mathbf{m}_z
+ \mu_{e^\prime g} \sin (\theta_{e^\prime g}) \mathbf{m}_x$, 
$\boldsymbol{\mu}_{fe} = \mu_{fe} \cos( \theta_{fe}) \mathbf{m}_z
+ \mu_{fe} \sin (\theta_{fe}) \mathbf{m}_x$
and 
$\boldsymbol{\mu}_{fe^\prime} = \mu_{fe^\prime} \cos( \theta_{fe^\prime}) \mathbf{m}_z
+ \mu_{fe^\prime} \sin (\theta_{fe^\prime}) \mathbf{m}_x$.
The angle between the different transition dipole moments is given by
\begin{equation}
\tan  (\theta_{\nu\mu}) = \frac{|\boldsymbol{\mu}_{eg} \times \boldsymbol{\mu}_{\nu\mu}| }{
\boldsymbol{\mu}_{eg} \cdot \boldsymbol{\mu}_{\nu\mu}}.
\end{equation}

The isotropically averaged signals can then be written in the compact form
\begin{equation} 
\mathbf{P}^{pq} = \mathsf{M}^{pq} \boldsymbol{\chi}^{pq}, \qquad p,q \in\{e,e^\prime\}
\end{equation}
with 
\begin{equation}
\begin{split}
\mathbf{P}^{ee}(T) &= \left[
\langle P^{e,e,e,e}_{\mathbf{z},\mathbf{z},\mathbf{z},\mathbf{z}} (0,T,0)\rangle_{\mathrm{iso}},
\langle P^{e,e,e,e^\prime}_{\mathbf{z},\mathbf{z},\mathbf{z},\mathbf{z}}(0,T,0)\rangle_{\mathrm{iso}},
\langle P^{e,e,e^\prime,e}_{\mathbf{z},\mathbf{z},\mathbf{z},\mathbf{z}}(0,T,0)\rangle_{\mathrm{iso}},
\langle P^{e,e,e^\prime,e^\prime}_{\mathbf{z},\mathbf{z},\mathbf{z},\mathbf{z}}(0,T,0)\rangle_{\mathrm{iso}}
\right],
\\
\mathbf{P}^{e^\prime e^\prime}(T) &= \left[
\langle P^{e^\prime,e^\prime,e,e}_{\mathbf{z},\mathbf{z},\mathbf{z},\mathbf{z}}(0,T,0)\rangle_{\mathrm{iso}},
\langle P^{e^\prime,e^\prime,e,e^\prime}_{\mathbf{z},\mathbf{z},\mathbf{z},\mathbf{z}}(0,T,0)\rangle_{\mathrm{iso}},
\langle P^{e^\prime,e^\prime,e^\prime,e}_{\mathbf{z},\mathbf{z},\mathbf{z},\mathbf{z}}(0,T,0)\rangle_{\mathrm{iso}},
\langle P^{e^\prime,e^\prime,e^\prime,e^\prime}_{\mathbf{z},\mathbf{z},\mathbf{z},\mathbf{z}}(0,T,0)\rangle_{\mathrm{iso}}
\right]
\\
\mathbf{P}^{ee^\prime}(T) &= \left[
\langle P^{e^\prime,e,e,e}_{\mathbf{z},\mathbf{z},\mathbf{z},\mathbf{z}} (0,T,0)\rangle_{\mathrm{iso}},
\langle P^{e^\prime,e,e,e^\prime}_{\mathbf{z},\mathbf{z},\mathbf{z},\mathbf{z}}(0,T,0)\rangle_{\mathrm{iso}},
\langle P^{e^\prime,e,e^\prime,e}_{\mathbf{z},\mathbf{z},\mathbf{z},\mathbf{z}}(0,T,0)\rangle_{\mathrm{iso}},
\langle P^{e^\prime,e,e^\prime,e^\prime}_{\mathbf{z},\mathbf{z},\mathbf{z},\mathbf{z}}(0,T,0)\rangle_{\mathrm{iso}} ,
\right. \\
&  \left. \hspace{0.6cm}
\langle P^{e,e^\prime,e,e}_{\mathbf{z},\mathbf{z},\mathbf{z},\mathbf{z}}(0,T,0)\rangle_{\mathrm{iso}},
\langle P^{e,e^\prime,e,e^\prime}_{\mathbf{z},\mathbf{z},\mathbf{z},\mathbf{z}}(0,T,0)\rangle_{\mathrm{iso}},
\langle P^{e,e^\prime,e^\prime,e}_{\mathbf{z},\mathbf{z},\mathbf{z},\mathbf{z}}(0,T,0)\rangle_{\mathrm{iso}},
\langle P^{e,e^\prime,e^\prime,e^\prime}_{\mathbf{z},\mathbf{z},\mathbf{z},\mathbf{z}}(0,T,0)\rangle_{\mathrm{iso}}
\right]
\end{split}
\end{equation}
and
\begin{equation}
\begin{split}
\boldsymbol{\chi}^{ee}(T) &= \left[
\chi^{e,e,e,e} (T),
\chi^{e^\prime,e^\prime,e,e}(T),
\Re \chi^{e,e^\prime,e,e}(T),
\Im \chi^{e,e^\prime,e,e}(T)
\right] ,
\\
\boldsymbol{\chi}^{e^\prime e^\prime}(T) &= \left[
 \chi^{e,e,e^\prime,e^\prime}(T),
\chi^{e^\prime,e^\prime,e^\prime,e^\prime}(T),
\Re \chi^{e,e^\prime,e^\prime,e^\prime}(T),
\Im \chi^{e,e^\prime,e^\prime,e^\prime}(T)
\right]
\\
\boldsymbol{\chi}^{ee^\prime}(T) &= \left[
\Re\chi^{e,e,e,e^\prime}(T),
\Re\chi^{e^\prime,e^\prime,e,e^\prime}(T),
\Re\chi^{e,e^\prime,e,e^\prime}(T),
\Re\chi^{e^\prime,e,e,e^\prime}(T),
\right. \\
&  \left. \hspace{0.6cm}
\Im\chi^{e,e,e,e^\prime}(T),
\Im\chi^{e^\prime,e^\prime,e,e^\prime}(T),
\Im\chi^{e,e^\prime,e,e^\prime}(T),
\Im\chi^{e^\prime,e,e,e^\prime}(T)
\right],
\end{split}
\end{equation}

\begin{equation}
\begin{split}
 \mathsf{M}^{ee}_{11} &=- \frac{2}{15} {\mu}_{ e g}^4,
\qquad
 \mathsf{M}^{ee}_{12} =  -\frac{1}{5} {\mu}_{ e g}^4 - (1-\Gamma)  \frac{1}{15} \left[ \cos(2\theta_{f e^\prime}) +2\right]
{\mu}_{ f e^\prime}^2 {\mu}_{ e g}^2
\\
\mathsf{M}^{ee}_{13} &=  \mathsf{M}^{ee}_{14} =  \mathsf{M}^{ee}_{21} =  \mathsf{M}^{ee}_{22} = 
\mathsf{M}^{ee}_{31} =  \mathsf{M}^{ee}_{32} =  \mathsf{M}^{ee}_{43} =  \mathsf{M}^{ee}_{44} = 0
\\
 \mathsf{M}^{ee}_{23} &= \mathsf{M}^{ee}_{33}= - \frac{1}{15} {\mu}_{ e g}^2 \left\{ 
(1-\Gamma) \left[ 3 \cos(\theta_{fe}) \cos(\theta_{fe^\prime}) + \sin(\theta_{fe}) \sin(\theta_{fe^\prime}) \right]
{\mu}_{ f e} {\mu}_{ f e^\prime} + 3 \cos(\theta_{e^\prime g})  {\mu}_{ eg} {\mu}_{ e^\prime g}   \right\}
\\
 \mathsf{M}^{ee}_{24} &= -\mathrm{i} \mathsf{M}^{ee}_{23},
\qquad
 \mathsf{M}^{ee}_{34} = \mathrm{i} \mathsf{M}^{ee}_{23},
\\
 \mathsf{M}^{ee}_{41} &= -\frac{1}{15} {\mu}_{ e g}^2
\left\{\left[ \cos(2\theta_{e^\prime g}) +2\right]{\mu}_{e^\prime g}^2  
+ (1-\Gamma)  \left[ \cos(2\theta_{f e}) +2\right]{\mu}_{f e}^2 \right\},
\qquad
 \mathsf{M}^{ee}_{42} = -\frac{2}{15} {\mu}_{ e g}^2 {\mu}_{e^\prime g}^2 
\left[ \cos(2\theta_{e^\prime g}) +2\right] 
\end{split}
\end{equation}

\begin{equation}
\begin{split}
\mathsf{M}^{e^\prime e^\prime}_{11} &=-\frac{2}{5} {\mu}_{ e g}^2 {\mu}_{e^\prime g}^2
\left[ \cos(2\theta_{e^\prime g}) +2\right],
\qquad
\mathsf{M}^{e^\prime e^\prime}_{12} =  -\frac{1}{5} {\mu}_{ e g}^2
\left\{\left[ \cos(2\theta_{e^\prime g}) +2\right]{\mu}_{e g}^2  
+ (1-\Gamma)  \left[ \cos(2\theta_{f e^\prime}-\theta_{e^\prime g}) +2\right]{\mu}_{f e^\prime}^2 \right\}
\\
\mathsf{M}^{e^\prime e^\prime}_{13} &= \mathsf{M}^{e^\prime e^\prime}_{14} = \mathsf{M}^{e^\prime e^\prime}_{21} = 
\mathsf{M}^{e^\prime e^\prime}_{22} = \mathsf{M}^{e^\prime e^\prime}_{31} = \mathsf{M}^{e^\prime e^\prime}_{32} = 
\mathsf{M}^{e^\prime e^\prime}_{43} = \mathsf{M}^{e^\prime e^\prime}_{44} = 0
\\
\mathsf{M}^{e^\prime e^\prime}_{23} & = \mathsf{M}^{e^\prime e^\prime}_{33} =- \frac{1}{15} {\mu}_{ e^\prime g}^2 \left\{ 
(1-\Gamma) \left[ 2 \cos(\theta_{fe}-\theta_{f e^\prime} ) + \cos(\theta_{fe} +\theta_{fe^\prime}-2\theta_{e^\prime g})\right]
{\mu}_{ f e} {\mu}_{ f e^\prime} + 3 \cos(\theta_{e^\prime g})  {\mu}_{ eg} {\mu}_{ e^\prime g}   \right\}
\\
\mathsf{M}^{e^\prime e^\prime}_{24} &= -\mathrm{i} \mathsf{M}^{e^\prime e^\prime}_{23},
\qquad
\mathsf{M}^{e^\prime e^\prime}_{34} = \mathrm{i} \mathsf{M}^{ee}_{23},
\\
\mathsf{M}^{e^\prime e^\prime}_{41} &= -\frac{1}{15} {\mu}_{ e^\prime g}^2
\left\{ 3  {\mu}_{ e^\prime g}^2+ (1-\Gamma) \left[ \cos(2\theta_{f e} - \theta_{e^\prime g}) +2\right]{\mu}_{f e}^2  \right\}
\\
\mathsf{M}^{e^\prime e^\prime}_{42} &= -\frac{2}{5} {\mu}_{ e^\prime g}^4 
\end{split}
\end{equation}

\begin{equation}
\begin{split}
\mathsf{M}^{e e^\prime}_{11} &=-\frac{2}{5} \cos(\theta_{e^\prime g}){\mu}_{ e g}^3 {\mu}_{e^\prime g},
\qquad
\mathsf{M}^{e e^\prime}_{12} =  -\frac{1}{15} {\mu}_{ e g}  {\mu}_{ e^\prime g}
\left\{ 3 \cos(\theta_{e^\prime g}) {\mu}_{e g}^2  
+ (1-\Gamma)  \left[ \cos(2\theta_{f e^\prime}-\theta_{e^\prime g}) +2 \cos(\theta_{e^\prime g}) \right]{\mu}_{f e^\prime}^2 \right\}
\\ 
\mathsf{M}^{e e^\prime}_{13} &= \mathsf{M}^{e e^\prime}_{14} = \mathsf{M}^{e e^\prime}_{17} = 
\mathsf{M}^{e e^\prime}_{18} 
\qquad
\mathsf{M}^{e e^\prime}_{16} = \mathrm{i} \mathsf{M}^{ee}_{11}
\qquad
\mathsf{M}^{e e^\prime}_{17} = \mathrm{i} \mathsf{M}^{ee}_{12}
\\
\mathsf{M}^{e e^\prime}_{21} &= \mathsf{M}^{e e^\prime}_{22} = 
\mathsf{M}^{e e^\prime}_{23} = \mathsf{M}^{e e^\prime}_{25} = \mathsf{M}^{e e^\prime}_{26} = 
\mathsf{M}^{e e^\prime}_{27} = 0
\\
\mathsf{M}^{e e^\prime}_{24} & = M^{e^\prime e^\prime}_{33} =- \frac{1}{15} {\mu}_{ e g}{\mu}_{ e^\prime g} \left\{ 
(1-\Gamma) \left[ 2 \cos(\theta_{fe}-\theta_{f e^\prime} -\theta_{e^\prime g}) + 2 \cos(\theta_{fe}) 
\cos(\theta_{fe^\prime}-\theta_{e^\prime g})\right]
{\mu}_{ f e} {\mu}_{ f e^\prime}  \right\}
\\
\mathsf{M}^{e e^\prime}_{28} &= -\mathrm{i} \mathsf{M}^{e^\prime e^\prime}_{24}
\\
\mathsf{M}^{e e^\prime}_{34} &= \mathrm{i} \mathsf{M}^{ee}_{23},
\qquad
\mathsf{M}^{e e^\prime}_{31} = \mathsf{M}^{e e^\prime}_{32} =
\mathsf{M}^{e e^\prime}_{34} = \mathsf{M}^{e e^\prime}_{35} = 
\mathsf{M}^{e e^\prime}_{36} = \mathsf{M}^{e e^\prime}_{38} =0,
\qquad
\mathsf{M}^{e e^\prime}_{37} = \mathrm{i} \mathsf{M}^{ee}_{23},
\\
\mathsf{M}^{e e^\prime}_{41} &= -\frac{1}{15} {\mu}_{ e g}  {\mu}_{ e^\prime g}
\left\{ 3 \cos(\theta_{e^\prime g}) {\mu}_{e^\prime g}^2  
+ (1-\Gamma)  \left[ \cos(2\theta_{f e}-\theta_{e^\prime g}) +2 \cos(\theta_{e^\prime g}) \right]{\mu}_{f e}^2 \right\}
\\
\mathsf{M}^{e e^\prime}_{42} &= -\frac{2}{5} \cos(\theta_{e^\prime g}){\mu}_{ e g} {\mu}_{e^\prime g}^3 
\\
\mathsf{M}^{e e^\prime}_{43} &= \mathsf{M}^{e e^\prime}_{44} =
\mathsf{M}^{e e^\prime}_{47} = \mathsf{M}^{e e^\prime}_{48} =0,
\qquad
\mathsf{M}^{e e^\prime}_{45} = \mathrm{i} \mathsf{M}^{ee}_{41},
\qquad
\mathsf{M}^{e e^\prime}_{46} = \mathrm{i} \mathsf{M}^{ee}_{42},
\\
\mathsf{M}^{e e^\prime}_{51} &= \mathsf{M}^{e e^\prime}_{11},
\qquad
\mathsf{M}^{e e^\prime}_{52} =  \mathsf{M}^{e e^\prime}_{12},
\qquad
\mathsf{M}^{e e^\prime}_{53} = \mathsf{M}^{e e^\prime}_{54} = \mathsf{M}^{e e^\prime}_{57} = 
\mathsf{M}^{e e^\prime}_{58} = 0
\qquad
\mathsf{M}^{e e^\prime}_{55} = - \mathrm{i} \mathsf{M}^{ee}_{11},
\qquad
\mathsf{M}^{e e^\prime}_{56} = - \mathrm{i} \mathsf{M}^{ee}_{12},
\\
\mathsf{M}^{e e^\prime}_{61} &= \mathsf{M}^{e e^\prime}_{62} = 
\mathsf{M}^{e e^\prime}_{64} = \mathsf{M}^{e e^\prime}_{65} = M^{e e^\prime}_{66} = 
\mathsf{M}^{e e^\prime}_{67} = \mathsf{M}^{e e^\prime}_{68} = 0
\qquad
\mathsf{M}^{e e^\prime}_{63}  = \mathsf{M}^{e^\prime e^\prime}_{24},
\qquad
\mathsf{M}^{e e^\prime}_{67}  = - \mathrm{i}  M^{e^\prime e^\prime}_{24} ,
\\
\mathsf{M}^{e e^\prime}_{71} &= \mathsf{M}^{e e^\prime}_{72} = \mathsf{M}^{e e^\prime}_{73} =
\mathsf{M}^{e e^\prime}_{75} = \mathsf{M}^{e e^\prime}_{76} = \mathsf{M}^{e e^\prime}_{77} = 0,
\qquad
\mathsf{M}^{e e^\prime}_{78}  = - \mathsf{M}^{e^\prime e^\prime}_{37},
\\
\mathsf{M}^{e e^\prime}_{81} &=\mathsf{M}^{e e^\prime}_{11},
\qquad
\mathsf{M}^{e e^\prime}_{82} =\mathsf{M}^{e e^\prime}_{42},
\\
\mathsf{M}^{e e^\prime}_{83} &= \mathsf{M}^{e e^\prime}_{84} =
\mathsf{M}^{e e^\prime}_{87} = \mathsf{M}^{e e^\prime}_{88} =0
\qquad
\mathsf{M}^{e e^\prime}_{85} = - \mathsf{M}^{ee}_{45}, 
\qquad 
\mathsf{M}^{e e^\prime}_{86} = - \mathsf{M}^{ee}_{86}
\end{split}
\end{equation}

In defining the vector $\boldsymbol{\chi}(T)$, besides its properties in Eq.~(\ref{equ:prop1})-(\ref{equ:prop3}), 
the following relations were instrumental,
$\chi_{ggee}(T) -1 = - \chi_{eeee}(T) - \chi_{e^\prime e^\prime ee}(T)$,
$\chi_{gge^\prime e^\prime}(T) -1 = - \chi_{ee e^\prime e^\prime }(T) 
- \chi_{e^\prime e^\prime e^\prime e^\prime }(T)$
and
$ \chi_{gge e^\prime}(T) = - \chi_{ee e e^\prime }(T) - \chi_{e^\prime e^\prime e e^\prime }(T)$.

\end{widetext}

\end{document}